\documentclass{article}

\usepackage[preprint, nonatbib]{nips_2018}

\usepackage{float}

\usepackage[square,numbers]{natbib}
\usepackage[utf8]{inputenc} % allow utf-8 input
\usepackage[T1]{fontenc}    % use 8-bit T1 fonts
\usepackage{hyperref}       % hyperlinks
\usepackage{url}            % simple URL typesetting
\usepackage{booktabs}       % professional-quality tables
\usepackage{amsfonts}       % blackboard math symbols
\usepackage{nicefrac}       % compact symbols for 1/2, etc.
\usepackage{microtype}      % microtypography
\usepackage{graphicx}
\usepackage{tabularx}
\usepackage{hyperref}
\usepackage{amsmath,amssymb}
\usepackage[ruled,longend,linesnumbered]{algorithm2e}

\title{Trend-Based SAC Beam Control Method with Zero-Shot in Superconducting Linear Accelerator}

\author{
	Xiaolong Chen \\
	\texttt{chenxiaolong@impcas.ac.cn} \\
	\And
	Xin Qi \\
	\texttt{qixin2002@impcas.ac.cn} \\
	\And
	Chunguang Su \\
	\texttt{suchunguang@impcas.ac.cn} \\
	\And
	Yuan He \thanks{Corresponding author.}\\ 
	\texttt{hey@impcas.ac.cn} \\
	\And
	Zhijun Wang \thanks{Corresponding author.}\\
	\texttt{wangzj@impcas.ac.cn} 
	\And
	Kunxiang Sun \\
	\texttt{sunkunxiang@impcas.ac.cn} \\
	\And
	Chao Jin \\
	\texttt{jinchao@impcas.ac.cn} \\
	\And
	Weilong Chen \\
	\texttt{chenweilong@impcas.ac.cn} \\
	\And
	Shuhui Liu \\
	\texttt{liush@impcas.ac.cn} \\
	\And
	Xiaoying Zhao \\
	\texttt{zxy@impcas.ac.cn} \\
	\And
	Duanyang Jia \\
	\texttt{jiaduanyang@impcas.ac.cn} \\
	\And
	Man Yi \\
	\texttt{yiman@impcas.ac.cn} 	
}

\begin{document}
	% \nipsfinalcopy is no longer used
	\maketitle
	
	\begin{abstract}
			The superconducting linear accelerator is a highly flexiable facility for modern scientific discoveries, necessitating weekly reconfiguration and tuning. Accordingly, minimizing setup time proves essential in affording users with ample experimental time. We propose a trend-based soft actor-critic(TBSAC) beam control method with strong robustness, allowing the agents to be trained in a simulated environment and applied to the real accelerator directly with zero-shot. To validate the effectiveness of our method, two different typical beam control tasks were performed on  China Accelerator Facility for Superheavy Elements (CAFe II) and a light particle injector(LPI) respectively. The orbit correction tasks were performed in three cryomodules in CAFe II seperately, the time required for tuning has been reduced to one-tenth of that needed by human experts, and the RMS values of the corrected orbit were all less than 1mm. The other transmission efficiency optimization task was conducted in the LPI, our agent successfully optimized the transmission efficiency of radio-frequency quadrupole(RFQ) to over $85\%$ within 2 minutes. The outcomes of these two experiments offer substantiation that our proposed TBSAC approach can efficiently and effectively accomplish beam commissioning tasks while upholding the same standard as skilled human experts. As such, our method exhibits potential for future applications in other accelerator commissioning fields.
	\end{abstract}

	\section{Introduction}
	Superconducting linear accelerator is a highly flexible scientific facility and has been widely used in scientific discovery. For example, the CAFe II can accelerate particles with a mass-to-charge ratio greater than or equal to $1/3$ in the energy range of $1.33MeV/u$ to $6 MeV/u$ \citep{cafe2019, CiADS,beamcommissioning2016}. In superconducting linear accelerator, achieving the required beam quality often involves adjusting dozens of magnet parameters by highly experienced engineers, which could consume a significant amount of valuable beam time. Machine learning (ML) and reinforcement learning (RL) algorithms have been demonstrated to successfully automate beam tuning in various applications. For example, \cite{2012ORBIT} employed an actor-critic framework to correct the trajectory of a storage ring in a simulated environment. And \cite{2017Using} used RL to train a controller for rapid switching between beam parameters in a free electron laser (FEL) . At FERMI, RL is employed to optimize accelerator control with satisfactory preliminary results reported \citep{TowardtheApplication2019,2020Basic,hirlaenderModelfree,osheaPolicyGradientMethods2020}. Furthermore, RL is also implemented to stabilize the operation of THz CSR in synchrotron light sources, overcoming instability limitations caused by bunch self-interaction \citep{2020Feedback}. These successes extend to the training of sample-efficient RL algorithms such as normalized advantage function (NAF) at CERN AWAKE electron line and the H-accelerator LINAC4 \citep{2020Sample}. By utilizing a highly efficient convergence algorithm, the period of online time required to perform orbit correction can be significantly reduced, achieving a minimum duration of 20 minutes. \cite{2021Application} and \cite{2021Orbit} trained controllers based on historical BPM data to relize online orbit correction. Methods that make use of historical data, such as SLAC's pulse energy tuning method based on the Bayesian optimization algorithm \cite{2020BayesianOptimization}. Most of these apporachs required the collection of historical data or even online training of machines. However, collecting and cleaning historical data is a challenging task, while online training requires valuable beam time. In this paper, we implemented a zero-shot tuning accelerator commissioning method -- without any data collection of real accelerator during training process, by using a trend-based soft actor-critic(SAC) algorithm\citep{softActorCritic2018}. The robustness of our TBSAC-based agents - trained in a simulated environment - was evaluated by performing two typical beam commissioning tasks: orbit correction in SC of CAFe II and beam transmission efficiency optimization in LEBT of the light particle injector. These agents were deployed on real accelerators without further modification. Our experimental findings indicate that, with respect to human standards, our agent can achieve beam commissioning tasks in a comparatively more efficient manner than humans. 
	
	We start with a brief introduction to RL and its potential for accelerator control in Section 2. In Section 3, we introduce our TBSAC method and its scheme. To evaluate the effectiveness of our approach, we present experimental results in Sections 4 and 5. Specifically, in Section 4, we demonstrated that our method achieved significant speed improvement in CAFe II compared to traditional orbit correction methods.In Section 5, we demonstrate how our approach enhances the transmission efficiency of RFQ in LPI by achieving a result comparable to human performance with greater efficiency. Finally, we summarize our findings and provide proposals for further research in Section 6.
	
	\section{Reinforcement learning}
	RL is a machine learning approach that empowers an agent to acquire an optimal behavior policy by interacting with the environment. The high operational costs of large accelerator facility necessitates the use of efficient beam commissioning methods to minimize beam tuning time, RL is commonly consider as one of the tuning method with great potential in accelerator domain.
	
	A standard paradigm of RL involves an agent optimizing its policy to maximize expected cumulative rewards by interacting with an environment. At each time step $t$, the agent selects actions $a \in A$ by observing the state $s \in S$ of the enviroment based on its policy $ \pi: S\rightarrow A$, and receives a reward $r$ and new states of the enviroment $s\prime$. The total return is defined as $R_t = \sum_{i=t}^{T}\gamma^{i-t}r(s_i,a_i)$ by summing up all the discounted rewards from the current time step until the end of the episode. The discount factor $\gamma$ balances the immediate and future rewards. To obtain an optimal policy, traditional RL maximizes the expected sum of rewards $\sum_{t}^{}\mathbb{E}_{(s_t,a_t) \sim\pi}[r(s_t,a_t)]$.  SAC is an efficient and stable model-free RL algorithm for continuous state and action spaces. And the aim of the agent is to maximize both the cumulative reward and the expected entropy of the policy $\pi$:
	$$ J(\pi) = \sum_{t=0}^{T}\mathbb{E}_{(s_t,a_t)\sim \pi}{[r(s_t,a_t)+ \alpha H(\pi(.|s_t))]}, $$
	where $\alpha$ is a temperature parameter to control the policy stochasticity $H$. The incorporation of entropy in SAC can help to alleviate the problem of premature convergence and facilitate effective exploration, which is especially useful when dealing with complex environments that contain a lot of noise and measurement errors. Due to unmeasurable errors that exist between simulated and real accelerator, as well as significant noise in the data collected from real accelerator, using the SAC algorithm to achieve zero-shot tuning in real accelerator is a good choice.

	\section{TBSAC beam control method with zero-shot}
\begin{figure*}
		\includegraphics[width=0.95\textwidth]{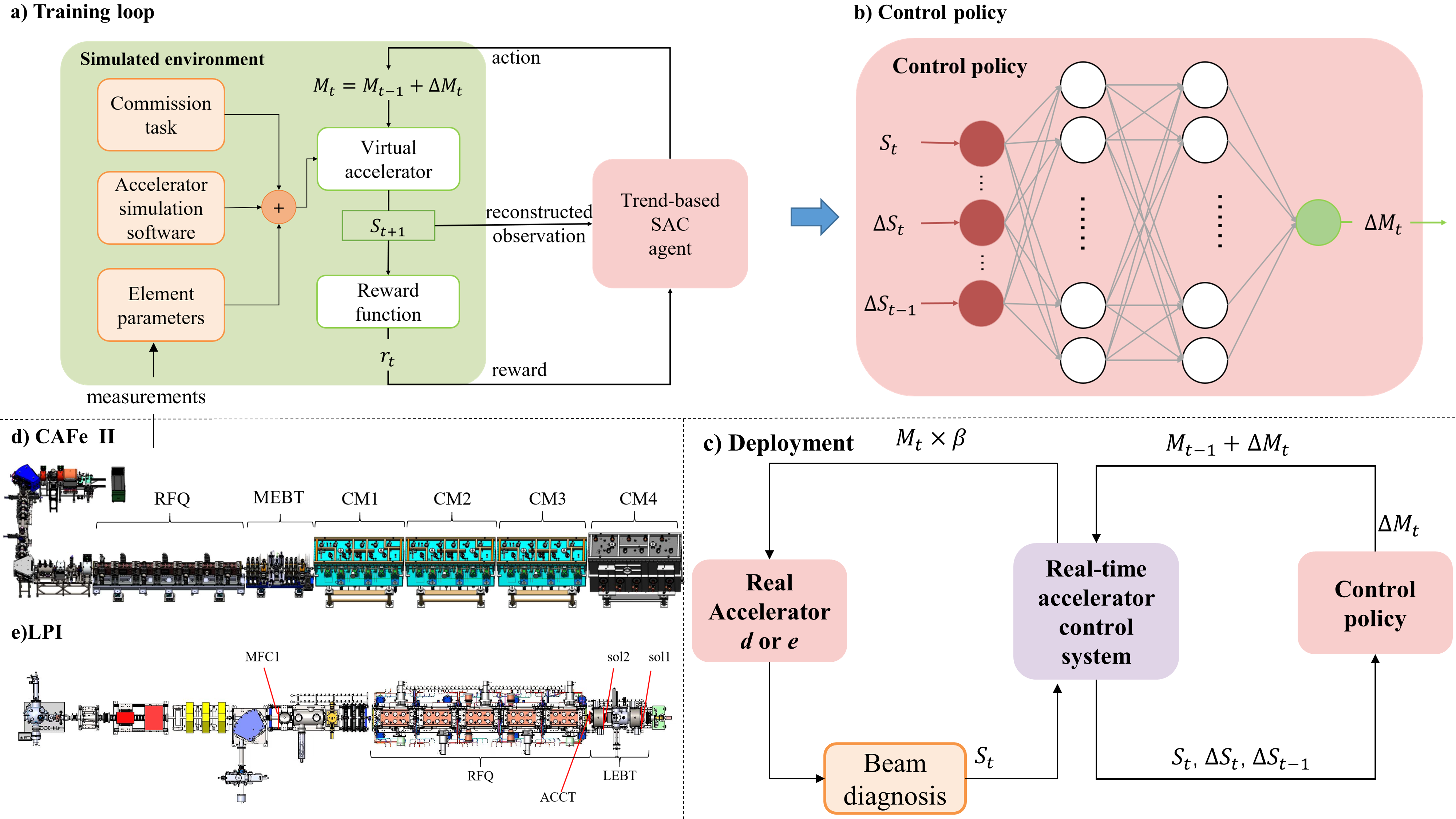}
		\caption{\label{fig:scheme} (a) Overview of the training loop and the structure of simulated environment. (b) The network structure of the trend-based SAC agent. The trend of the observed variable is incorporated into the observation space. (c) The trained control policy can be applied on the real accelerator directly. (d) The overview of CAFe II. (e) The overview of LPI.}
	\end{figure*}

	A zero-shot\citep{zeroshot} tuning method can be defined as follows: an agent trained in a simulated environment can be directly applied to a real environment without further training or adjustment. To address the zero-shot tuning problem in beam commissioning task, we adopted a methodology which is shown in Fig. ~\ref{fig:scheme}. 
	
		\IncMargin{1em}
	\begin{algorithm}[H]
		\renewcommand{\algorithmcfname}{Method}
		\caption{Training proccess of TBSAC}\label{algorithm}
		\SetKwData{Left}{left}\SetKwData{This}{this}\SetKwData{Up}{up}
		\SetKwFunction{Union}{Union}\SetKwFunction{FindCompress}{FindCompress}
		\SetKwInOut{Input}{Input}\SetKwInOut{Output}{Output}
		Initialize replay buffer $D$ to capacity $N$ \; 
		Initialize neuro network with random weights \;
		\For{ $t=1,T$}{
			\For{each episode}{
				Reset random elements errors\;
				$OB_{zero} \leftarrow SimulatedEnv(0)$\;
				$OB_{inital}  \leftarrow SimulatedEnv(var)$, where $var$ is a variable\;
				$\Delta OB_{inital}  \leftarrow OB_{inital}-OB_{zero}$\;
				$\Delta OB_{zero} \leftarrow 0$\;
				Inital state $S_{inital} = [OB_{inital}, \Delta OB_{inital}, \Delta OB_{zero}]$\;
				\eIf{$t < random steps$}{
					Sample $\Delta M_t$ from action space randomly\; 	
				}{
					Determine $\Delta M_t$ by the observation   $S_{t} = [OB_t, \Delta OB_t, \Delta OB_{t-1}]$\;
				}
				$S_{t+1},r_t,dw \leftarrow SimulatedEnv(M_{t-1}+ \Delta M_t)$
				
				Save $(S_t, \Delta M_t, r_t, S_{t+1},dw)$ into replaybuffer $D$\;
				\If{$t > random steps$}{
					Sample a batch of data from $D$ and update the policy\;
				}
				
			}
		}
	\end{algorithm} 
	
	Firstly, it is imperative to create a simulation environment that reflects the basic physical properties of the actual accelerator as accurately as possible. The structure of our simulated environment is shown in the green part of Fig. ~\ref{fig:scheme}(a). The virtual accelerator is developed based on a accelerator simulation software, Tracewin\citep{tracewin} for us, and the input and output of it are determined by the specific commissioning tasks. Measured data of the elements in real accelerator is merged into the accelerator simulation software. Secondly, although the measured parameters are incorporated into the simulation environment, there still exist unmeasurable errors between the real and simulated environments. And this presents a significant challenge to zero-shot tuning. In order to address this issue, we have developed a trend-based agent that leverages the experience of human engineers. The agent observes changes in the variable to be optimized and determines the necessary adjustments based on their observed trends. By incorperating the trend of the observed variables, the agent can learn the invariant properties of the system during the training process. Finally, taking the trends-based reward as a part of the reward function can provide guidance for the agent during the training process. To clarify, Method ~\ref{algorithm} shows a train process of our TBSAC agent, where $OB_t$ stands for the value of the target optimization variable read by beam diagnostic equipment at time $t$, and $M$ represents the current of the magnet that needs to be regulated.
	
	In this paper, beam control tasks are abstracted as Markov decision processes(MDP), and TBSAC is employed to address these optimal problems. To verify the effectiveness of our method, two kinds of typical beam control tasks -- orbit correction task in the SC of CAFe II and transmission efficency optimization of LPI  were performed. The implementation of TBSAC is based on \citep{drlcode}, and the source code \footnote{https://github.com/ElivisC/TBSAC, we will release the source code as soon as possible.} of our implementation is available online.
	
		\section{Orbit correction in CAFe II}
	
\begin{figure}[h]
		\includegraphics[width=0.95\textwidth]{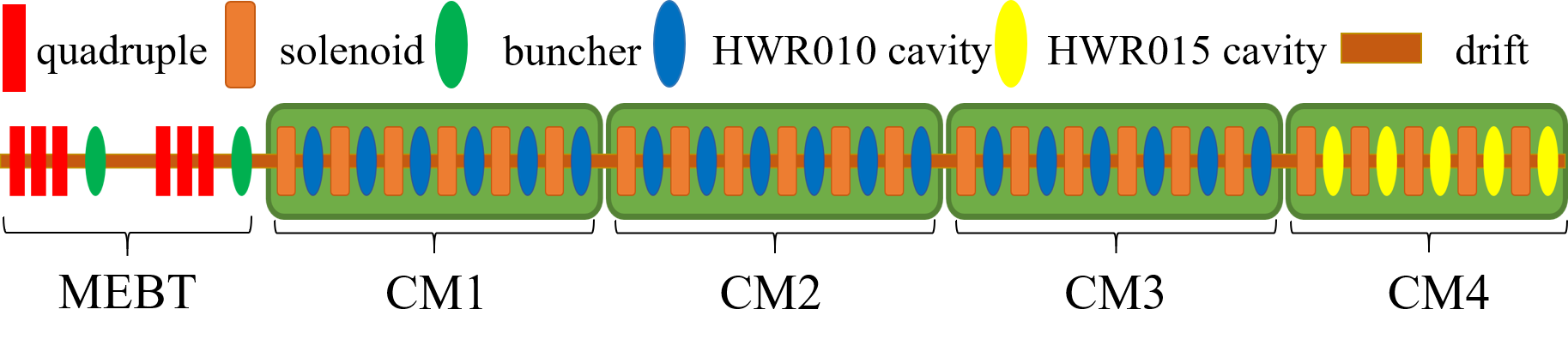}
		\caption{\label{fig:cafe_detail} The detail structure of the MEBT section and the SC in CAFe II. The MEBT section consists of six quadruple magnets and two bunchers, spanning a length of 2.6975 meters. The superconducting section contains four cryomodule(CM), the first three modules each contain six HWR010 cavities and the last one contains five HWR019 cavities\citep{beamcommissioning2016}.}
	\end{figure}

	CAFe II is a superconducting linear accelerator which is mainly used to synthesis of superheavy elements. The overview of CAFe II is shown in Fig. ~\ref{fig:scheme}(d) and the detail structure of MEBT section and SC of CAFe II are shown in Fig. ~\ref{fig:cafe_detail}.  In CAFe II, the machine needs to be reconfigured at least once a week depends on the varying energy demands. Due to the nonlinear property caused by the fringe field effects\citep{fringefields2015} and space charge forces\citep{spacecharge2014,onthe2022,differentable2023}, the tuning process is diffcult for human and the required time for correction can vary from 30 to 150 minutes, depending on the operator's proficiency. Reducing the time required for the tuning processes as much as possible is of great significance. For CAFe II, we chose the first three modules to complete the orbit correction task in SC cause of there only 2 groups of BPMs are worked in CM4.

	The goal of orbit correction task is to minimize the BPM readings $B$ as close to 0 as possible by adjusting the magnet strength $M$.  To ensure consistency in physical properties between virtual and real environments, measurement data from the real accelerator is merged into the configuration files of the software. In our case, we measured the polarity of the solenoids, the polarity of the corrector magnets, and the installation errors (confusion in the X and Y directions) individually. And merged them into the configuration file of Tracewin to create a simulated environment. 
	
	The initial state of each episode is a set of different installation errors of solenoids, which means that the agent can be trained to deal orbit correction task with various errors. Specifically, a random error of $\pm3mm$ is added to each solenoid in the superconducting section, and a $\pm0.6mm$ error is added to the last quadrupole magnet of MEBT to widen the distribution of SC's entry. The solenoid error causes an approximate deviation of $\pm9mm$ for the beam center, while the quadrupole magnet error results in an approximately  $\pm3mm$ deviation of the beam before entering the SC. At each time step $t$, the agent makes decisions based on observation $B_t$ and adjusts the strength of correctors from $M_{t-1}$ to $M_t$ by tuning an increment $\Delta M$, then the $M_t$ will be taken as the input of the accelerator simulation software to obtain the new state $B_{t+1}$. By executing this loop, an episode will be terminated when the RMS of BPM readings is less than $0.5mm$ or when the correction steps exceed 50. In order to reduce the dimensionality of the parameter space, we connected each of the three crymodules to the exit of MEBT section instead of restore the full structure of SC completely, and three independent environments for CM1, CM2 and CM3 were created respectively. In each module, 5 groups of correctors and 5 groups of BPMs are employed to correct the orbit of the beam in the $x$ and $y$ directions. A python wrapper based on TraceWin is developed in simulated environment and the environments are created by using  \cite{gym} gym framework. To deploying the agent on real accelerator, an online environment is built by using both OpenAI gym and pyepics.
	
\begin{figure}
		\includegraphics[width=0.9\textwidth]{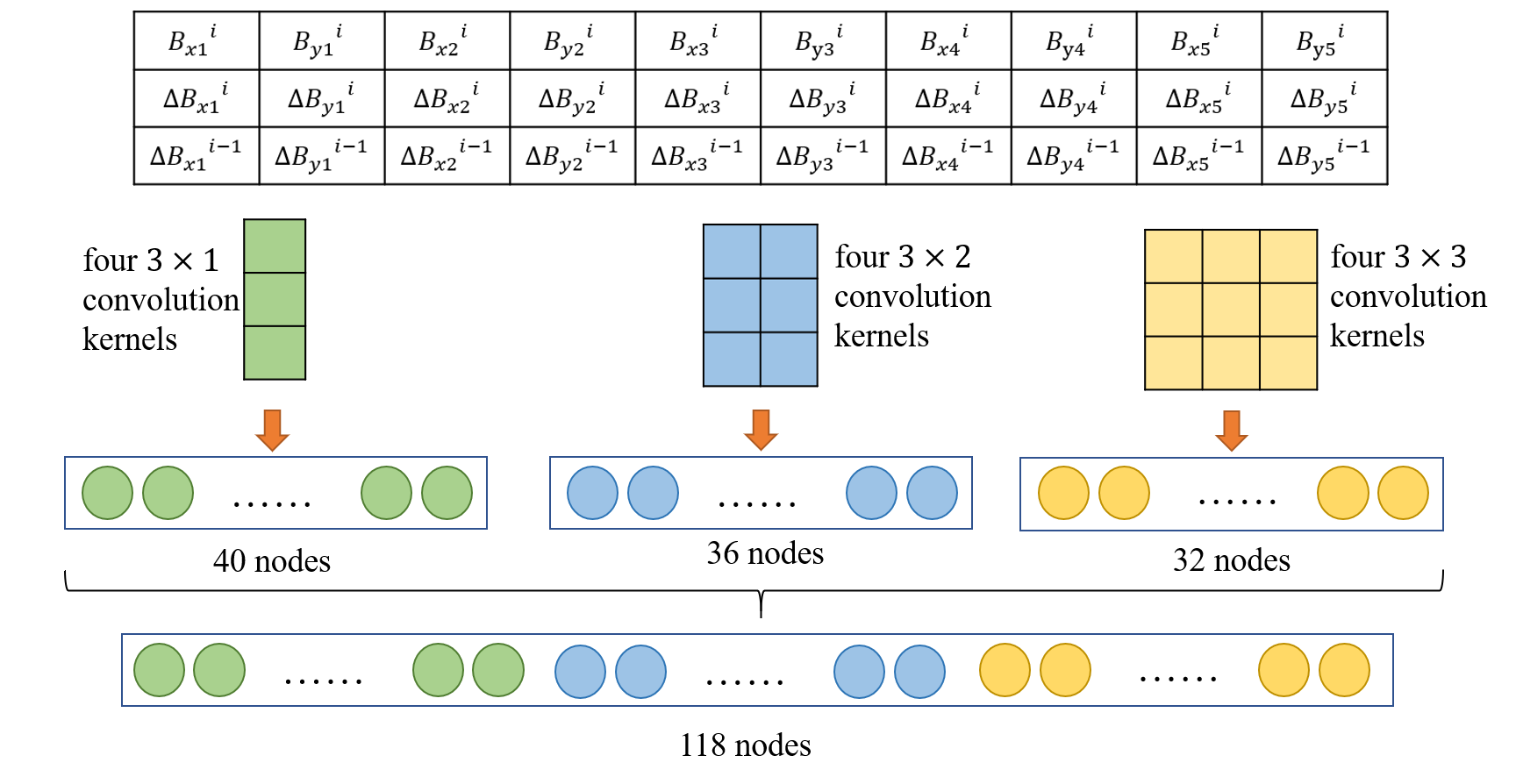}
		\caption{\label{fig:cnn_structure} The special CNN layer in the actor. }
	\end{figure}

	To incorporate the trend of BPM readings, the input of the actor is taken as $[B_t,\Delta B_t, \Delta B_{t-1}]$ or $[B_t,\Delta B_t, \Delta B_{t-1}]^T$ , depends on the different network structure of the actor, such as DNN or convolutional neuron network(CNN) in this paper and according to our experiments, both of the two structures are effective. In this paper, we present experimental results based on the CNN network architecture. Fig. ~\ref{fig:cnn_structure} illustrates the special CNN layer we have inserted between the input and first hidden layer of the actor .  In order to provide a more clear trend guidance, our reward function is set as 
	$$ r_t = r_{rms} + \alpha r_{trend},$$
	$$ \Delta B_{t\_abs} = \left | B_t \right | - \left | B_{t-1} \right |,$$
	$$  r_{trend} = \sum_{i=0}^{2N}{\begin{cases} 
			1/N, & \Delta B_{t\_abs}[i] < 0\\
			-4/N, & \Delta B_{t\_abs}[i] \geq 0
	\end{cases} } ,$$
	
	where $r_{rms}$ is the negative RMS value of BPM readings between $B_t$ and zero, $ alpha $ is the factor of the importance of trend and N is the total number of all the BPMs.

\begin{figure}
		\includegraphics[width=0.8\textwidth]{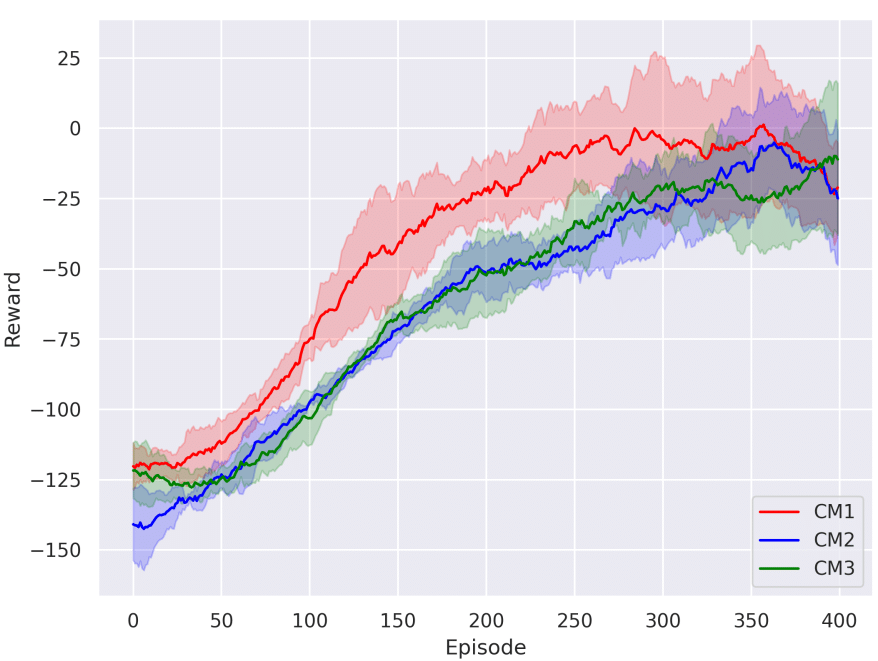}
		\caption{\label{fig:train_curves} The smoothed train curves of the three crymodules. }
	\end{figure}

\begin{figure}
		\includegraphics[width=0.95\textwidth]{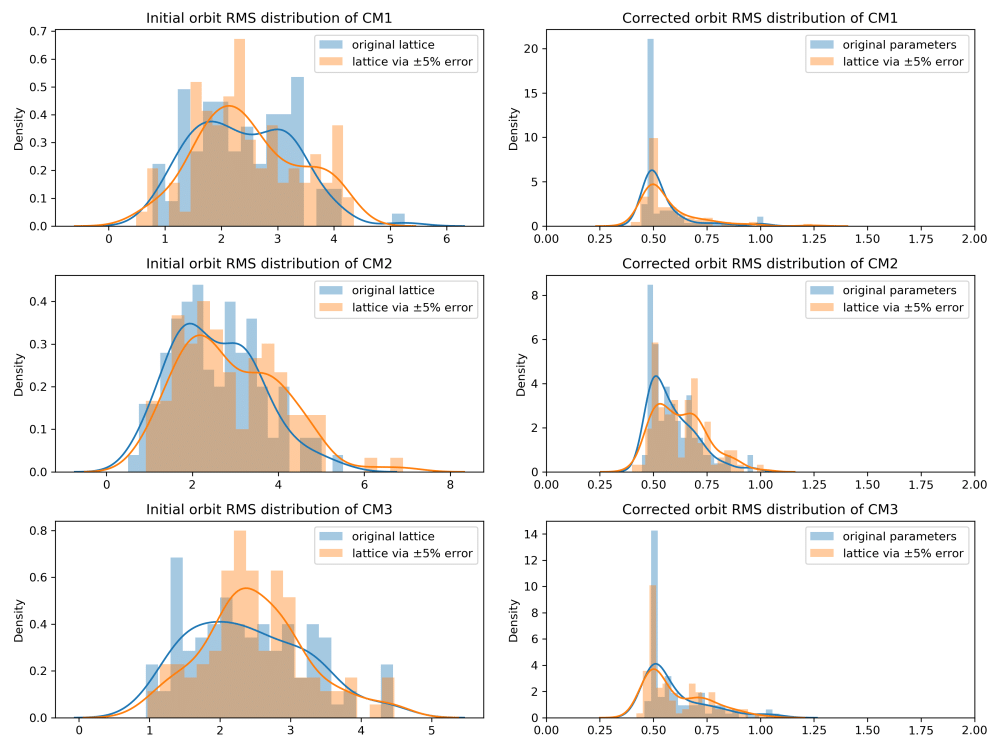}
		\caption{\label{fig:simulation} The statistical results of 200 simulation experiments for CM1-CM3 in the orbit correction task. Among them, 100 experiments were conducted without errors, while the other 100 experiments added errors of $\pm 5\% $.}
	\end{figure}

\begin{figure}
		\includegraphics[width=0.95\textwidth]{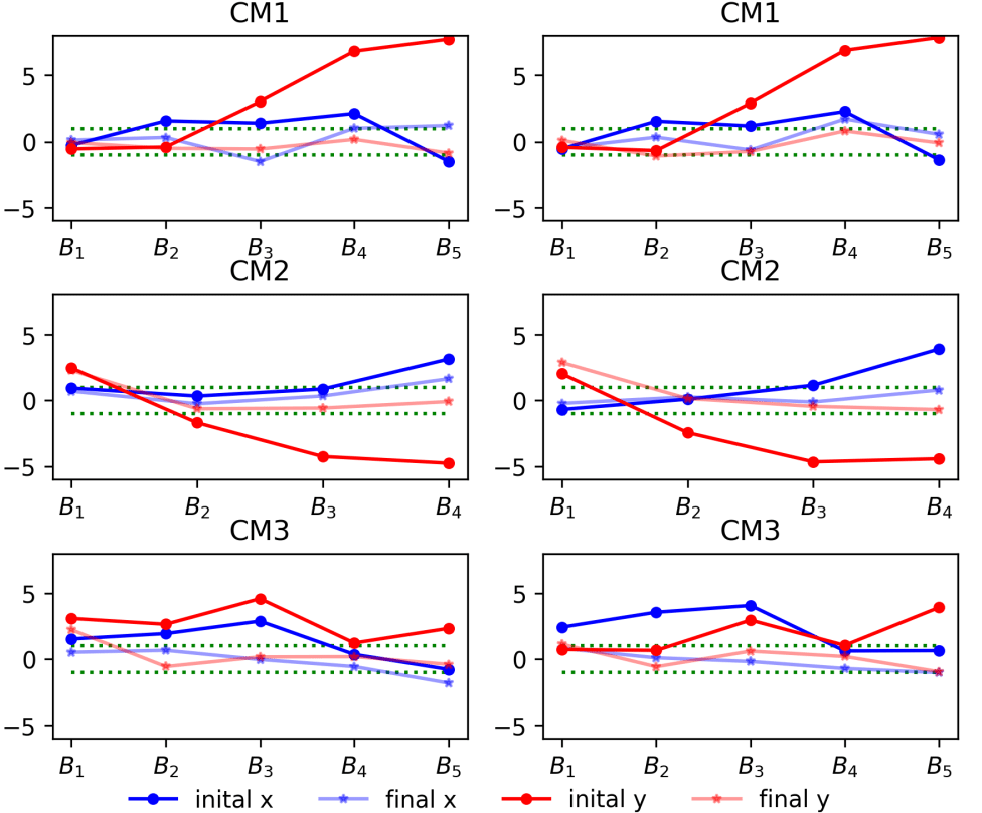}
		\caption{\label{fig:caferesult} The orbit correction results of CM1-CM3 in CAFe II. The green dashed line represents a BPM reading of -1 or 1.}
	\end{figure}

	We conducted experiments using 5 different random seeds for each module to verify the stablity of our method. And the training process of the agents in the simulated environment is shown in Fig. ~\ref{fig:train_curves}.  The weights of 10000 time steps were chosen to evaluate the effectiveness of our agents. We firstly tested the agents on simulated environment, each module for 200 experiments, 100 experiments for the original lattice and the other 100 experiments for the lattice with a random error. Fig. ~\ref{fig:simulation} provides a comparison of the RMS data distribution of the orbits before and after correction in 100 orbit correction tasks for the three modules. The results indicate that the agent can effectively correct at least $97\%$ of the initial orbits to an RMS of less than 1mm. To verify the robustness of the agents, we add a $\pm 5\%$ random error on the strength of solenoids. And The results shows that the agent can still correct at least $98\%$ of the initial orbits to an RMS of less than 1mm.  According to the results, we can see that there was a slight decline in the proportion of RMS less than 0.5mm, and this is acceptable for orbit correction tasks.
	Two online correction experiments performed in CAFe II. We applied the agents trained by  $^{40}Ar^{12+}$ in simulated environment to correct the orbits of CM1, CM2, and CM3 in CAFe II. CM1 and CM3 each have five groups of correctors and BPMs, while CM2 has four groups of correctors and BPMs due to damage of its last BPM. The left column in Fig. ~\ref{fig:caferesult} shows the corrected results of the three modules for the $40Ar{12+}$ beam with an energy of 230 MeV provided by CAFe II. The agents corrected CM1, CM2, and CM3 orbits in 29, 33, and 23 steps, respectively. The agents accomplished the tasks in approximately 3 minutes (1 minute per module), reducing the initial RMS value of all modules to below 1mm. This represents at least a 10-fold improvement in efficiency compared to human commissioning engineers, greatly enhancing the efficiency of beam commissioning. Another experiment was conducted with the lattice adjusted according to the energy requirement of 222 MeV, during which the maximum deviation of solenoid current intensity reached $5\%$. In this condition, we applied the previous agents without any modification to do the tasks. As a result, these agents successfully corrected the orbit RMS to below 1mm on CM1-3 within 17, 22, and 15 steps respectively. And the results are shown in the right column of Fig. ~\ref{fig:caferesult}. These experiments demonstrate that due to the strong robustness of our agents, they can overcome the difference between the simulated environment and real accelerator, even with a $\pm 5\%$ error at least according current experiments.
	
	\label{title_page}

	\section{Transmission effiency optimization of RFQ in LPI}
	LPI is a small accelerator experimental machine used to provide proton beams and Fig.\ref{fig:scheme} (e) shows the structure of the LPI. Transmission effiency of RFQ  is an essential process in the tuning process \citep{BeamMatching}. Engineers observe the beam currents strength at the exits of LEBT and RFQ to adjust the currents of two solenoids and two groups of correction magnets, thus enhancing the transmission efficiency of LEBT and RFQ. The goal of this task is to enhance the transmission efficiency of RFQ as much as possible while meeting the requirements for the beam current strength of LEBT exit. In this task, we first trained a high-accuracy surrogate model \citep{2021Realtime} to enable fast transport of the particle beam . Then we created an interactive environment based on this surrogate model and employed TBSAC for policy learning. Finally, we validated the effectiveness of the model in both Tracewin and the LPI.
	
	Simulating 10,000 particles for beam loss is time-consuming, hindering RL training and hyper-parameter tuning\citep{drlthatmatters2017} due to the single iteration taking a minimum of 5 minutes. A surrogate model was implemented using a neural network comprised of an input layer with 2 nodes, 3 hidden layers with 16, 32, 16 nodes respectively, and an output layer with 2 nodes. After training, a model was chosen with an R-squared value exceeding $98\%$ to create a simulated environment. 
	
	At the start of each episode, the initial values of the solenoid are fixed, and the beam current of LEBT and the transmission efficiency of RFQ are both situated in a position far from the optimal solution. In simulation, termination of an episode occurs when the LEBT exit beam current is above $70\%$ and RFQ transmission efficiency is over $95\%$ . In the actual operating conditions of the experiment, the termination conditions were determined based on the experience of the engineering team. Specifically, the beam current at the exit of the LEBT was required to be higher than $30\%$ compared to the ion source exit beam intensity(due to the presence of impurity particles), and the transmission efficiency of RFQ had to exceed $85\%$. Additionally, tuning process of the episode will also end if tuning steps exceed 50. Consistent with the orbit correction task, we integrated the trend in transmission efficiency changes in RFQ into the reward function, while the beam current in LEBT serves as a physical boundary constraint.  
	
	\begin{figure}
		\includegraphics[width=0.95\textwidth]{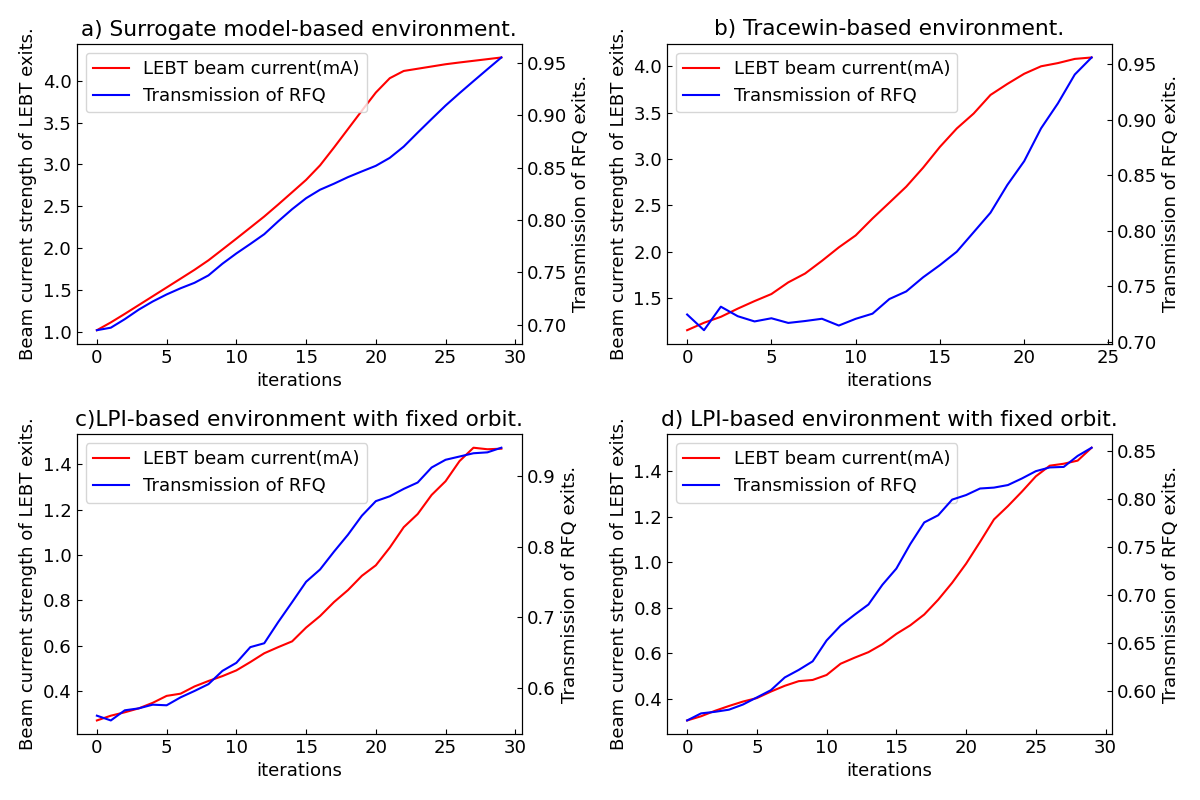}
		\caption{\label{fig:transmission} Two experiments were conducted to optimize the transmission efficiency in LEBT of LPI for different fixed orbits.}
	\end{figure}
	
	The actor's network architecture was simplified due to small input and output dimensions, comprising an input layer that receives the state trend $[S_t, \Delta S_t, \Delta S_{t-1}]$ and two dense hidden layers, each with 64 node. The output consists of the adjustment amounts for two solenoid. The training process was completed in approximately 5 minutes, and we selected the weights after 9000 steps for the verification of policy effectiveness in both surrogate model-based and Tracewin-based environment. 
	
	The results in Fig.~\ref{fig:transmission}(a)(b) indicate that the agent can optimize the transmission efficiency of LEBT and RFQ to the designated targets in these two simulated environments. Two online transmission efficency optimization tasks were performed in LPI with two specific different trajectories. The agent trained in the surrogate model-based environment were applied on the LPI directly without any modification. The results in Fig. ~\ref{fig:transmission}(c)(d) indicate that our agent can optimize the transmission efficiency of RFQ to 85\% and 93\%, respectively, within 2 minutes, while the beam currents strength of LEBT exits is $1.48mA$ and $1.52mA$, which is basically the same as the results obtained by engineer tuning. These experiments indicate that our TBSAC agents have enough robustness to applied in a real accelerator with zero-shot. The proposed TBSAC can significantly enhance beam tuning efficiency, rivaling that of human commissioning engineers. 
	
	\section{Conclusion}
	In this paper, we proposed a trend-based SAC algorithm to address the beam control problems with zero-shot tuning in the real accelerator. The trend of observed variables is incorporated into the observation space and a reward function is designed as a guidance. Two typical beam control problems are used to verify the robustness of our apporach. And the results show that in the orbit correction task, we have reduced the tuning time from 30-120 minutes to within 3 minutes, greatly improving the operational efficiency of the accelerator. In beam transmission efficiency optimization tasks, our agent achieved results comparable to those of human experts in optimizing transmission efficiency within 2 minutes. All of these results demonstrate that TBSAC can overcome the difference between the simulated environment and the real accelerator, and can effectively improve the accelerator commissioning efficiency. The approach presented in this study holds promising potential for generalization to various other areas of accelerator beam control.
	
	\section*{Acknowledgements}
	This work was supported by the National Natural Science Foundation of China (Grant No.U22A20261) and Large Research Infrastructures "China initiative Accelerator Driven System" (Grant No.2017-000052-75-01-000590).

	\bibliography{sc.bib}

\end{document}